\documentclass[twoside,a4paper,12pt]{article}
\usepackage[utf8]{inputenc}
\usepackage[T1]{fontenc}
\usepackage{textcomp}
\usepackage{lmodern}
\usepackage[english]{babel}
\usepackage{amsmath}
\usepackage{amsfonts}
\usepackage{amsthm}
\usepackage{amssymb}
\usepackage{layout}
\usepackage[top=2.5cm, bottom=2.5cm, left=2.5cm, right=2.5cm]{geometry}
\usepackage{graphicx}
\usepackage{color}
\usepackage{xcolor}
\usepackage[extra]{tipa}
\usepackage{ulem}
\usepackage{fancyhdr}
\usepackage{wrapfig}
\usepackage{epstopdf}
\usepackage{float}
\usepackage{authblk}
\usepackage{hyperref}
\usepackage{authblk}
\usepackage[toc,page]{appendix}
\usepackage{chemist} 
\usepackage[version=3]{mhchem} 
\usepackage{subfig} 
\usepackage{caption}

\newcommand{\qmax}{q_{\textrm{max}}}
\newcommand{\mumax}{\mu_{\textrm{max}}}

\title{Termination mechanisms of Turing patterns in growing systems}

\author[1,2]{Gabriel Morgado}
\author[2]{Laurence Signon}
\author[1,3]{Bogdan Nowakowski}
\author[2]{Annie Lemarchand}
\affil[1]{Institute of Physical Chemistry, Polish Academy of Sciences, Kasprzaka 44/52, 01-224 Warsaw, Poland}
\affil[2]{Sorbonne Université, Centre National de la Recherche Scientifique (CNRS), Laboratoire de Physique Théorique de la Matière Condensée (LPTMC), 4 place Jussieu, case courrier 121, 75252 Paris CEDEX 05, France}
\affil[3]{Warsaw University of Life Sciences (SGGW), Department of  Physics, 02-776 Warsaw, Poland}
\begin{document}
\maketitle
Corresponding author: Annie Lemarchand, E-mail: anle@lptmc.jussieu.fr

\begin{abstract}
The question of the termination of a periodic spatial structure of Turing type
in a growing system is addressed in a chemical engineering perspective and a biomimetic approach.
The effects of the dynamical parameters on the stability and the wavelength of the structure are analytically studied
and used to propose experimental conditions for which a Turing pattern stops by itself with a decreasing wavelength.
The proposed mechanism is successfully checked by the numerical integration of the equations governing the dynamics of the activator and the inhibitor.
We conclude that a local increase of the concentration of the reservoir which controls the injection rate of the inhibitor 
into the system can be used to achieve the appropriate termination of a Turing pattern. 
\end{abstract}
\newpage
\baselineskip=24pt
\section{Introduction}
During embryonic development, segmented structures of the body such as the spine and the digits are formed by the production of repeated elements.
Since the seminal work of Turing \cite{turing} accounting for the formation of biological pattern in the framework of reaction-diffusion models,
experimental evidences of Turing structures have been given in chemistry \cite{dekepper,sagues,epstein} and biology \cite{murray,raspopovic}.
Recent years have witnessed a growing number of papers where reaction-diffusion principles are proposed to model the formation
of biological periodic spatial structures \cite{maini,marcon,jcp139,economou,green,diambra,pre16}.
Following Turing, a two-component chemical system composed of an autocatalytically-produced activator by consumption of an inhibitor that diffuses
faster may produce periodic spatial structures such as stripes in one-dimensional (1D) systems and hexagons in 2D. In other words, a Turing pattern emerges by local
self-activation and lateral inhibition \cite{meinhardt}.
The concepts developed to model living systems provide a source of inspiration in chemical engineering \cite{grzybowski,jcpturing,mikhailov,taylor,steinbock,estevez,pre2018}.
However, standard models of Turing patterns generate structures in infinite systems and 
the question of the termination of a striped structure in a finite system arises in a perspective of biomimetism in material science.
Specifically, it is desirable to find experimentally achievable conditions creating a finite-size structure, whose growth stops by itself with decreasing oscillation amplitude
and respects the decrease of the   
wavelength during the termination process.
To this aim, we use an as simple as possible reaction-diffusion model \cite{epl} admitting a Turing structure developing behind a propagating wave front
 and examine the effect of all parameters on both the stability and the wavelength of the structure \cite{murray,pre2018}.
We already used a stochastic approach to a Turing pattern \cite{epl} and showed that, contrary to intuition,
the internal fluctuations may have a constructive effect and stabilize the structure outside the domain of stability
predicted by a deterministic description. Here, we adopt a macroscopic approach.
Our goal is to select suitable conditions from this systematic approach and to propose termination mechanisms compatible with processing in chemical engineering.\\

The paper is organized as follows. In section 2, a reaction-diffusion model involving local activation and long-range inhibition is presented. An analytical stability condition and the wavelength
expression of the Turing pattern are given. The influence of the parameters of the model on the stability and the wavelength of the pattern are studied in section 3. The analysis of the results
leads to the selection of a user-friendly termination mechanism in the framework of chemical engineering. The analytical predictions regarding stability and wavelength are compared to numerical results for the chosen mechanism.
Section 4 contains conclusions. The possibility that the different mechanisms exhibited could be found as termination scenarios in morphogenesis is raised.

\section{Model}
We consider the following reaction mechanism inspired from the Schnakenberg model \cite{schnakenberg} and the Gray-Scott model \cite{gray} 
\begin{eqnarray}
\label{r1}
\rm{A} &\begin{array}{c} 
\mbox{\footnotesize $k_1$} \\ \longrightarrow \\ \mbox{}
\end{array}& \rm{R_1}\\
\label{r2}
\rm{2A + B} &\begin{array}{c} 
\mbox{\footnotesize $k_2$} \\ \longrightarrow \\ \mbox{}
\end{array}& \rm{3A}\\
\label{r3}
\rm{B} & \hspace{1cm} \begin{array}{c}
\mbox{\footnotesize $k$}_{3}\\ \rightleftharpoons\\ \mbox{\footnotesize $k$}_{-3}
        \end{array} \hspace{1cm} &\rm{R_2}
\end{eqnarray}
where R$_1$ and R$_2$ are reservoirs. The concentrations $R_1$ and $R_2$ of the reservoirs are assumed to be constant in time. 
The reaction given in Eq.~(\ref{r2}) autocatalytically produces species A and consumes species B.
Due to the ability of accelerating its own production, species A is called an activator
whereas species B, which is consumed by the same process, is called an inhibitor.
The macroscopic dynamics of the system
is governed by two partial differential equations \cite{jcp139,epl} 
\begin{eqnarray}
\label{eqA}
\frac{\partial A}{\partial t} &=& -k_1 A+k_2 A^2B +D_A \frac{\partial^2 A}{\partial x^2} \\
\label{eqB}
\frac{\partial B}{\partial t} &=& k_{-3}R_2-k_3 B-k_2 A^2B +D_B \frac{\partial^2 B}{\partial x^2}
\end{eqnarray}
for the concentrations $A$ and $B$ of the activator and the inhibitor supposed to have different diffusion coefficients $D_A$ and $D_B$.
For appropriate rate constant values, such that
\begin{equation}
\label{delta}
\Delta=(k_{-3}R_2)^2-4k_1^2k_3/k_2 \geq 0
\end{equation}
the system admits two steady states
$(A_0=0,B_0=k_{-3}R_2/k_3)$ and
\begin{eqnarray}
\label{AT}
A_T&=&\frac{k_{-3}R_2+\sqrt{\Delta}}{2k_1}\\
\label{BT}
B_T&=&\frac{k_{-3}R_2-\sqrt{\Delta}}{2k_3}
\end{eqnarray}
that are stable with respect to homogeneous perturbations.
A linear stability analysis of Eqs.~(\ref{eqA},\ref{eqB}) reveals that the steady state $(A_T,B_T)$ can be destabilized
by inhomogeneous perturbations \cite{murray,sagues,jcp139,epl}.
\begin{figure}
\begin{center}
\includegraphics[width=10cm]{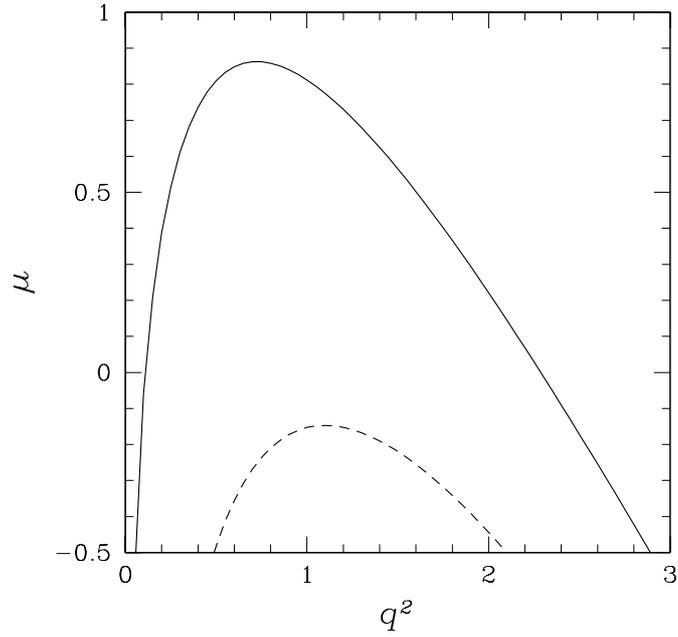}
\caption{ Largest eigenvalue $\mu$ of the linear operator $M$ versus square of Fourier mode $q^2$.
Solid line: $k_{-3}R_2=8.76$. Dashed line: $k_{-3}R_2=10$.
Other parameter values: $k_1=2.92$, $k_2=1$, $k_3=2.19$, $D_A=1$, $D_B=10$.
}
\end{center}
\end{figure}
The Fourier transforms $A_q(t)=\int_{-\infty}^\infty A(x,t)e^{-iqx}dx$ and 
$B_q(t)=\int_{-\infty}^\infty B(x,t)e^{-iqx}dx$, where $q$ is the Fourier mode, are introduced. In the Fourier space, the linear stability operator $M$ is given by:
\begin{eqnarray}
M = 
\begin{pmatrix}
k_1 - D_Aq^2 & k_2A_T^2 \\
-2k_1        & -\frac{k_2k_{-3}R_2}{k_1}A_T-D_Bq^2
\end{pmatrix}\label{M}
\end{eqnarray}
The eigenvalues of the matrix $M$ obey the characteristic equation $\mu^2+\alpha \mu +\beta=0$, with $\alpha=k_1-\frac{k_2k_{-3}R_2}{k_1}A_T-(D_A+D_B)q^2$
and $\beta=2k_1^2A_T/B_T-(k_1-D_Aq^2)(k_{-3}R_2/B_T+D_Bq^2)$.
The Turing structure develops if the largest eigenvalue
\begin{eqnarray}
\label{mu}
\mu&=&\dfrac{1}{2}\left(k_1-\frac{k_2k_{-3}R_2}{k_1}A_T-(D_A+D_B)q^2+\right. \nonumber\\
& &\left.\sqrt{\left(k_1+\frac{k_2k_{-3}R_2}{k_1}A_T+(D_B-D_A)q^2\right)^2-8k_1k_2A_T^2}\right)
\end{eqnarray}
is real and positive \cite{murray,sagues}.
Indeed, a system of differential equations, linearized around a homogeneous steady state, is easily solved by diagonalizing the
linear operator. Then, the solution is a linear combination of eigenmodes which exponentially depend on time
according to the corresponding eigenvalues. A term associated with a real, positive eigenvalue grows in time, leading to trajectories
in the concentration space that move away from the fixed point \cite{murray}. In the studied system, the destabilization of the
steady state occurs in favor of a Turing pattern.
Equation (\ref{mu}) imposes conditions on the parameter values.
In particular, the diffusion coefficient $D_B$ of the inhibitor B must be sufficiently larger than the diffusion coefficient $D_A$ of the activator A:
The destabilization of the homogeneous steady state $(A_T,B_T)$ requires local self-activation, ensured by the autocatalytic production of the
activator through the reaction given in Eq. (\ref{r2}), as well as long-range inhibition, due to the larger diffusion coefficient of the inhibitor.  
The mode $\qmax$, which maximizes the eigenvalue $\mu$, is the most unstable Fourier mode:
\begin{equation}
\label{qmax}
\qmax=\sqrt{\frac{A_T(D_A+D_B)\sqrt{2k_1k_2D_A/D_B}-k_1-k_2k_{-3}R_2A_T/k_1}{D_B-D_A}}
\end{equation}

\begin{figure}
\begin{center}
\includegraphics[width=10cm]{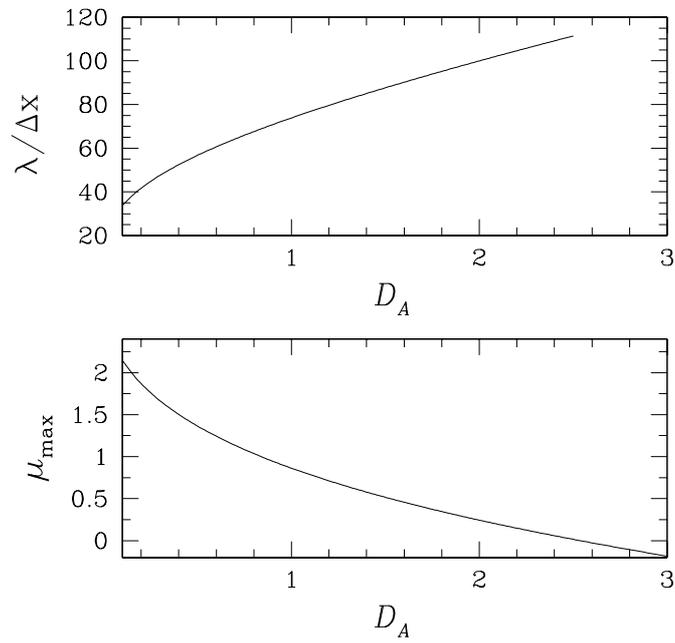}
\caption{Top: Scaled wavelength $\lambda/\Delta x$ of Turing pattern versus diffusion coefficient $D_A$ of species A.
Bottom: Maximum value $\mumax$ of the largest eigenvalue of the linear operator $M$ versus $D_A$.
Parameter values: $k_1=2.92$, $k_2=1$, $k_3=2.19$, $k_{-3}R_2=8.76$, $D_B=10$, $\Delta x=0.1$.
}
\end{center}
\end{figure}

In order to account for the termination of the Turing pattern in a growing system, including the fact that the structure ends with a gradually shorter spatial oscillation, we need to
find conditions for which the structure tends to lose its stability while its wavelength decreases.
The wavelength of the periodic structure is given by:
\begin{equation}
\label{lam}
\lambda=\dfrac{2\pi}{\qmax}
\end{equation}
Turing structure becomes unstable as the value of largest eigenvalue vanishes for the mode $\qmax$ associated with the maximum of $\mu$:
\begin{eqnarray}
\label{mumax}
\mumax&=&\dfrac{1}{2}\left(k_1-\frac{k_2k_{-3}R_2}{k_1}A_T-(D_A+D_B)\qmax^2+ \right .\nonumber\\
& &\left . \sqrt{\left(k_1+\frac{k_2k_{-3}R_2}{k_1}A_T+(D_B-D_A)\qmax^2\right)^2-8k_1k_2A_T^2}\right)
\end{eqnarray}
with $\qmax$ given in Eq. (\ref{qmax}).
Figure 1 illustrates the behavior of $\mumax$ for parameter values associated with a stable Turing pattern with
$\mumax >0$. It is also shown that it is sufficient to increase the value of
$k_{-3}R_2$ to shift the curve $\mu(q^2)$ down and lose the stability of the Turing pattern.\\

\begin{figure}
\begin{center}
\includegraphics[width=10cm]{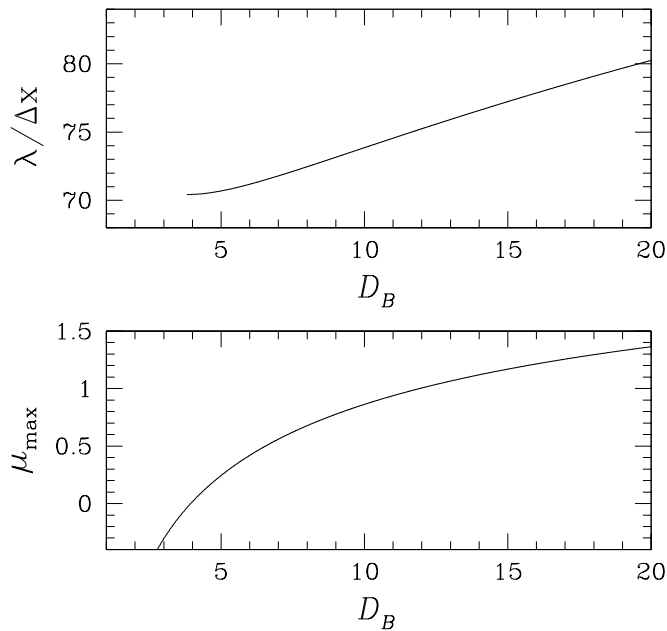}
\caption{Top: Scaled wavelength $\lambda/\Delta x$ of Turing pattern versus diffusion coefficient $D_B$ of species B.
Bottom: Maximum value $\mumax$ of the largest eigenvalue of the linear operator $M$ versus $D_B$.
Parameter values: $k_1=2.92$, $k_2=1$, $k_3=2.19$, $k_{-3}R_2=8.76$, $D_A=1$, $\Delta x=0.1$.
}
\end{center}
\end{figure}

In the next section we investigate the behavior of $\lambda$ and $\mumax$ as each parameter controlling dynamics varies.
Specifically, we aim at identifying diffusion coefficients or rate constants whose variation leads both to a decrease of the wavelength and a destabilization
of the Turing structure, i.e. negative values for the maximum of the eigenvalue.

\section{Results}
The concentration $R_2$ of the inhibitor reservoir is first assumed to be homogeneous. 
Figures 2 and 3 show the variations of the wavelength $\lambda$ and the maximum value $\mumax$ of the eigenvalue with one of the diffusion coefficients, the other parameters being constant.
The results are deduced from Eqs. (\ref{qmax}) and (\ref{lam}) for $\lambda$ and Eq. (\ref{mumax}) for $\mumax$, the expressions of the steady state $(A_T,B_T)$
being given in Eqs. (\ref{AT}) and (\ref{BT}). To facilitate the comparison with the numerical integration of Eqs. (\ref{eqA}) and (\ref{eqB}) that will be performed in the following, the 
wavelength is given in number of spatial cells of length $\Delta x=0.1$.
As shown in Fig. 2, the decrease in the maximum value $\mumax$ of the eigenvalue as the diffusion coefficient $D_A$ of species A increases is accompanied 
by an increase of the wavelength $\lambda$: The loss of stability of the Turing structure occurs with an increase of the spatial period.
We conclude that a variation of the diffusion coefficient $D_A$ cannot be argued as a justification of the termination process.
The behavior with respect to the diffusion coefficient $D_B$ of species B is different. 
The simultaneous loss of stability of the structure and the decrease of the wavelength are observed in Fig. 3 as $D_B$ decreases: The diffusion coefficient $D_B$ of species B
can be considered as a suitable parameter in the search for a termination model.\\

\begin{figure}
\begin{center}
\includegraphics[width=10cm]{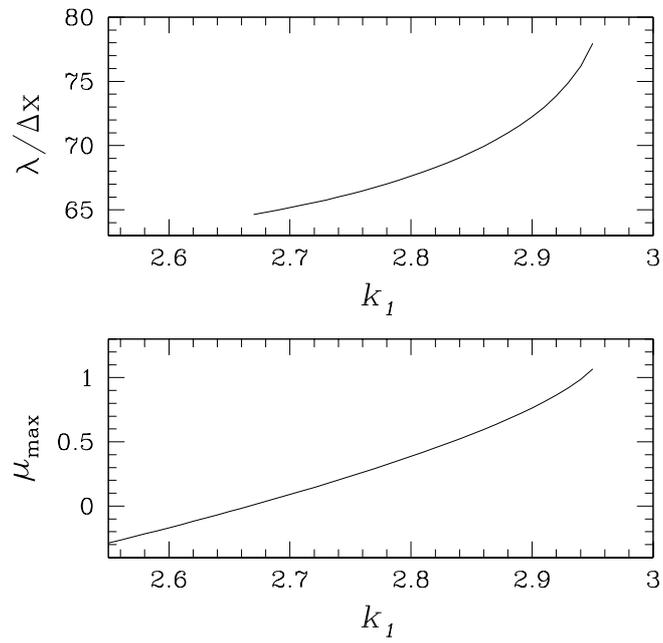}
\caption{Top: Scaled wavelength $\lambda/\Delta x$ of Turing pattern versus rate constant $k_1$ of the chemical reaction given in Eq. (\ref{r1}).
Bottom: Maximum value $\mumax$ of the largest eigenvalue of the linear operator $M$ versus $k_1$.
Parameter values: $k_2=1$, $k_3=2.19$, $k_{-3}R_2=8.76$, $D_A=1$, $D_B=10$, $\Delta x=0.1$.
}
\end{center}
\end{figure}

Figures 4-7 show the variations of the wavelength $\lambda$ and the maximum value $\mumax$ of the eigenvalue with rate constants.
The variations of $\lambda$ are given in a bounded interval of rate constant values, in which the Turing pattern is stable. At one of the endpoints of the interval,
the eigenvalue $\mumax$ vanishes and at the other endpoint, the condition of existence of the steady state $(A_T,B_T)$ given in Eq. (\ref{delta}) is no longer satisfied.
The two desired behaviors, i.e. the decrease of both $\lambda$ and $\mumax$, are observed as $k_1$ decreases, $k_2$ increases, $k_3$ decreases, and $k_{-3}R_2$ increases.
For an assumed homogeneous concentration $R_2$ of the reservoir, the variations of $\lambda$ and $\mumax$ with $R_2$ are analogous to 
the variations with $k_{-3}R_2$.
According to the chemical reaction given in Eq. (\ref{r1}), decreasing the rate constant $k_1$ tends to increase the concentration of species A.
Following Eq. (\ref{r2}), increasing the rate constant $k_2$ of the autocatalytic step tends to increase the concentration of species A and
decrease the concentration of species B.
This last result seems to be inconsistent with the consequences drawn from the decrease in $k_3$ or the increase in $k_{-3}R_2$, which result in increasing the concentration of species B
according to Eq. (\ref{r3}).
However, we already stated that increasing $B$ through soliciting the reservoir $R_2$ results in consuming species B faster by the autocatalytic step given 
in Eq. (\ref{r2}) \cite{jcp139,ctp}.
In particular, we observed that introducing a local source of species B leads to the nonintuitive local decrease of B concentration.
Hence, all the variations of the rate constants that lead to a loss of stability of the Turing pattern are eventually associated with an increase of A concentration and a decrease of B concentration.\\

\begin{figure}
\begin{center}
\includegraphics[width=10cm]{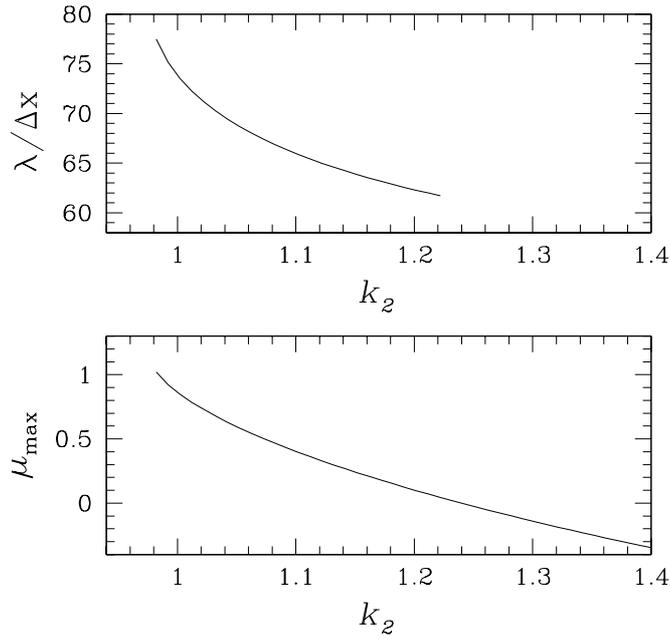}
\caption{Top: Scaled wavelength $\lambda/\Delta x$ of Turing pattern versus rate constant $k_2$ of the chemical reaction given in Eq. (\ref{r2}).
Bottom: Maximum value $\mumax$ of the largest eigenvalue of the linear operator $M$ versus $k_2$.
Parameter values: $k_1=2.92$, $k_3=2.19$, $k_{-3}R_2=8.76$, $D_A=1$, $D_B=10$, $\Delta x=0.1$.
}
\end{center}
\end{figure}

The diffusion coefficients and the rate constants characterize dynamics and are intrinsic to the reaction-diffusion system.
Nevertheless, it is always possible to imagine spatial variations of the dynamical parameters. 
Well-chosen variations of the diffusion coefficient $D_B$ of the inhibitor and each of the four rate constants of the chemical mechanism could be a priori used to build a termination model.
In the framework of the application to developmental biology, steric hindrance and molecular crowding in the tail of an embryo may be invoked to justify 
the decrease of the diffusion coefficients.
In chemical engineering, a local increase of temperature could be used to induce a local increase of the rate constants.
However, local increase of confinement or temperature is susceptible to simultaneously affect several dynamical parameters \cite{pre16,pre2018,minton1,minton2,weinstein,xie,kekenes,lipniacki}.
Whereas a decrease of $D_B$ is desired to destabilize the Turing pattern while decreasing its wavelength, a simultaneous decrease of $D_A$ would be detrimental.
Similarly, an increase of $k_2$ and $k_{-3}$ due to temperature increase could be satisfying but the joint decrease of $k_1$ and $k_3$
could blur the effect on the Turing structure.
The simplest way to imagine the control of a targeted parameter leading to the desired behavior is to impose well-chosen spatial variations 
of the reservoir concentration $R_2$. Indeed, the product $k_{-3}R_2$ plays the role of an apparent rate constant for the backward reaction given in Eq. (\ref{r3}) that can be fixed at will
in chemical engineering in the case of a single dynamical system with uniquely defined intrinsic parameters. \\

\begin{figure}
\begin{center}
\includegraphics[width=10cm]{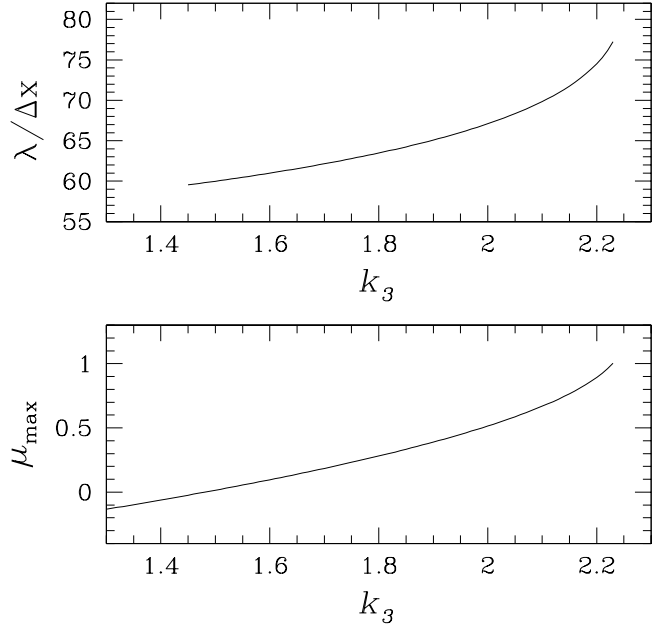}
\caption{Top: Scaled wavelength $\lambda/\Delta x$ of Turing pattern versus rate constant $k_3$ of the forward chemical reaction given in Eq. (\ref{r3}).
Bottom: Maximum value $\mumax$ of the largest eigenvalue of the linear operator $M$ versus $k_3$.
Parameter values: $k_1=2.92$, $k_2=1$, $k_{-3}R_2=8.76$, $D_A=1$, $D_B=10$, $\Delta x=0.1$.
}
\end{center}
\end{figure}

According to Fig. 7, increasing $R_2$ tends to destabilize the Turing pattern and decrease its wavelength. We examine if the results deduced from a stability analysis
can be used in a dynamical approach.
The results of the numerical integration of Eqs. (\ref{eqA}) and (\ref{eqB}) for a homogeneous concentration $R_2$ and a piecewise linear profile are given in Fig. 8.
The initial condition is a step function between the steady state $(A_T,B_T)$ in the first $10$ cells on the left and the steady state $(A_0,B_0)$ in the remaining cells.
The initial total number of cells is set at $n_0=610$.
Introducing the cell label $i=x/\Delta x$, where $\Delta x$ is the cell length, and the discrete time $s=t/\Delta t$, where $\Delta t$ is the time step, we choose:
\begin{eqnarray}
A(i,s=0)=A_T,\qquad B(i,s=0)=B_T, \qquad {\rm for} \qquad 1 \leq i \leq 10\\
A(i,s=0)=A_0,\qquad B(s,s=0)=B_0, \qquad {\rm for} \qquad 11 \leq i  \leq n_0 
\end{eqnarray}
To account for the growth of the system and simultaneously
avoid boundary effects that may alter the wavelength of the structure \cite{jcpturing}, 
we impose a fixed boundary on the left and a free growing end on the right \cite{jcp139,epl,ctp}.
For parameter values for which the steady state $(A_T,B_T)$ is unstable with respect to inhomogeneous perturbations, a Turing pattern 
develops after the passage of a chemical wave front. 
More precisely, according to Eqs. (\ref{eqA}) and (\ref{eqB}) and due to the no-flux boundary conditions applied on the left boundary, the concentrations in the first cell obey
\begin{eqnarray}
A(1,s+1)&=&A(1,s)-k_1\Delta tA(1,s)+k_2\Delta tA(1,s)^2B(1,s)+ \nonumber\\
        & &D_A\frac{\Delta t}{(\Delta x)^2}(A(2,s)-A(1,s))\\ 
B(1,s+1)&=&B(1,s)+k_{-3}R_2\Delta t-k_3\Delta tB(1,s)-k_2\Delta tA(1,s)^2B(1,s)+\nonumber \\
        & &D_B\frac{\Delta t}{(\Delta x)^2}(B(2,s)-B(1,s)) 
\end{eqnarray}
so that both $A(1,s)$ and $B(1,s)$ are extremum of the Turing pattern in the first spatial cell $i=1$.\\

Spatial cells are added to the right end of the system at the front speed to counterbalance the progression of the wave front and mimic system growth:
At all the discrete times $s$ for which the concentration $B(n-600,s)$ of species B in the $n-600$ cell becomes smaller than $0.99B_0$, 
the total number $n$ of cells is increased by 1. Provided that the front propagates at a speed smaller than
$\Delta x/\Delta t$, this protocol ensures that a layer of about $600$ cells remains in the stationary state $(A_0,B_0)$ on the right of the system, so that the propagation
of the front is not significantly affected by the finite size of the system.
To draw Fig. 8b, we have chosen the parameter values given in the caption of Figs. 1 and imposed $k_{-3}=8.76$ for the following spatial profile for the concentration $R_2$ of the inhibitor reservoir:
\begin{eqnarray}
R_2&=&1, \qquad {\rm for} \, 1 \leq i < 500\\
R_2&=&2.83 \times 10^{-4}i + 0.858,   \qquad {\rm for} \, 500 \leq i < 1000\\
R_2&=&1.14, \qquad {\rm for} \, 1000 \leq i 
\end{eqnarray}
\begin{figure}
\begin{center}
\includegraphics[width=10cm]{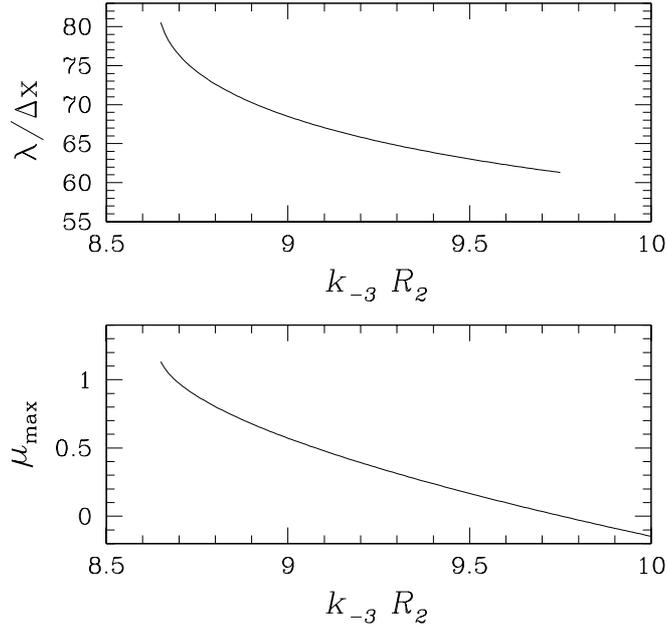}
\caption{Top: Scaled wavelength $\lambda/\Delta x$ of Turing pattern versus rate constant $k_{-3}R_2$ of the backward chemical reaction given in Eq. (\ref{r3}).
Bottom: Maximum value $\mumax$ of the largest eigenvalue of the linear operator $M$ versus $k_{-3}R_2$.
Parameter values: $k_1=2.92$, $k_2=1$, $k_3=2.19$, $D_A=1$, $D_B=10$, $\Delta x=0.1$.
}
\end{center}
\end{figure}
The simulation is stopped at time $t_{end}$ for which the wave front has passed cell $i = 1000$. 
It is worth noting that the Turing pattern is unchanged for larger values of the final integration time. Then, only the position of the
concentration gradients associated with the traveling wave evolve in time but the Turing structure has stopped growing and remains in a steady state with a fixed number
of wavelengths. As desired, the increase of the concentration $R_2$ leads to the termination of the Turing structure. \\

As illustrated in Fig. 7, the Turing structure is expected to be stable in the range $1 \leq i < 500$ for which $k_{-3}R_2=8.76$
and unstable in the range $i \geq 1000$ for which $k_{-3}R_2=10$.
More precisely, according to Eq. (\ref{mumax}),
the maximum of the eigenvalue $\mumax$ vanishes for $k_{-3}R_2=9.75$, i.e. $R_2=1.11$ for $k_{-3}=8.76$, which occurs in spatial cell $i=900$. 
Hence, the Turing pattern is predicted to be stable in the range $0 \leq i < 900$ and unstable beyond this domain.
The results shown in Fig. 8b confirm the analytical predictions.
The amplitude of the spatial oscillations 
decreases between $i \simeq 500$ and $i \simeq 1000$. The system is in a steady state in the range $ 1000 \leq i <1500$.\\
The increase of $R_2$ not only destabilizes the Turing structure but also modifies the steady state values and the propagation speed of the wave front.
The comparison between Figs. 8a and 8b shows that, as $R_2$ increases, the wave front propagates faster, $A_T$ increases, $B_T$ decreases and $B_0$ increases. 
As a consequence of the variation of $A_T$ and $B_T$, the oscillations of A and B concentrations are not symmetrical in the range $500 \leq i < 900$.
The decrease of the wavelength predicted in Fig. 7 is more difficult to check by a qualitative analysis. 
Using the numerical results illustrated in Fig. 8b, we evaluate the local wavelength by computing the number of cells between two minima of the A concentration profile.
The results are given in Fig. 9 and compared to the analytical prediction deduced from Eqs. (\ref{qmax}) and (\ref{lam}). 
The agreement between the numerical and analytical results is very satisfying in the range $600 \leq i < 900$. Oscillations of very small amplitude are observed in Fig. 8b
in the range $900 \leq i < 1000$, proving that a very damped Turing structure remains in a small area where instability was predicted.
The wavelength of the structure in the range $1 \leq i < 500$ is slightly affected by the increase of $R_2$ from cell $i = 500$ but the deviation from
the analytical prediction is only $2.5$ percent. 
This small difference is related to the linear approximation used in wavelength evaluation that neglects non-linear terms that may be more important for large structures.
Interestingly, the wavelength is sensitively decreased in the expected area in which 
the concentration of the reservoir R$_2$ increases: As shown in Fig. 9, the wavelength is reduced from 72 spatial cells to less than 61, before the structure disappears.
We conclude that an increase in the concentration of the reservoir R$_2$ related to the inhibitor $B$ is sufficient to account for the
destabilization of the Turing pattern associated with a decrease of the wavelength.
As anticipated by the results given in Fig. 7, according to which an increase of $R_2$ decreases the wavelength $\lambda$ and leads to a negative eigenvalue $\mu_{\max}$
around $(A_T,B_T)$, we suggest that an appropriate spatial variation of $R_2$ can be used in chemical engineering to stabilize the homogeneous steady state
and induce a termination of the Turing pattern in a growing system.
\begin{figure}
\begin{center}
{(a)}\includegraphics[width=8cm]{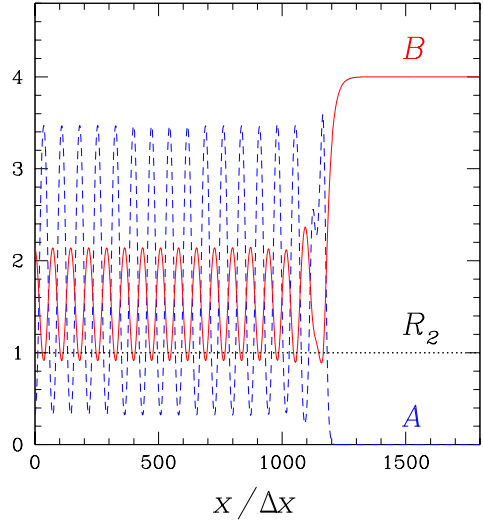}\\
{(b)}\includegraphics[width=8cm]{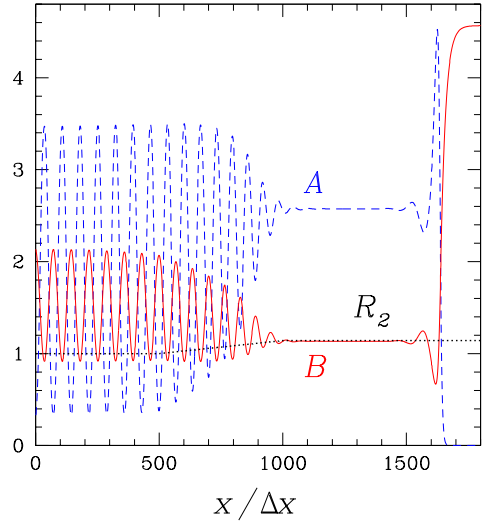}
\caption{Spatial profiles deduced from the numerical integration of Eqs. (\ref{eqA}) and (\ref{eqB}) for
$k_1=2.92$, $k_2=1$, $k_3=2.19$, $k_{-3}=8.76$, $D_A=1$, $D_B=10$,
$\Delta t=10^{-4}$, $t_{end}/\Delta t=2000000$, $\Delta x=0.1$.
Black dotted line: Imposed concentration $R_2$ of the reservoir. (a): Homogeneous concentration $R_2=1$, (b): Piecewise linear $R_2$ profile.
Blue dashed line: Concentration of species A versus cell label $x/\Delta x$.
Red solid line: Concentration of species B versus cell label $x/\Delta x$.
}
\end{center}
\end{figure}

\section{Conclusion}
\begin{figure}
\begin{center}
\includegraphics[width=10cm]{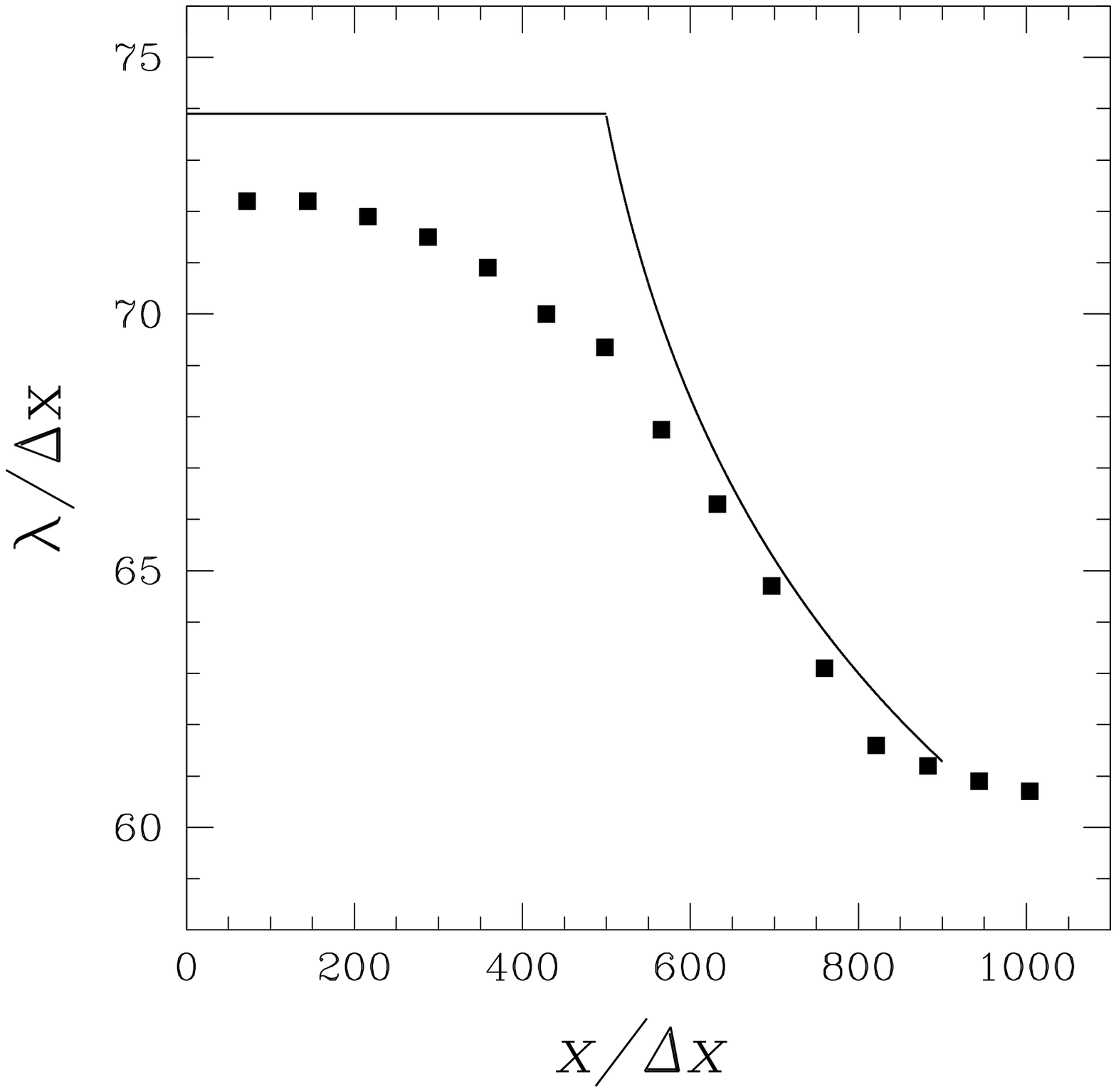}
\caption{Spatial variation of the scaled wavelength $\lambda/\Delta x$ of Turing pattern versus cell label $x/\Delta x$
for the parameter values given in the caption of Fig. 8b.
Symbols: Results deduced from the numerical integration of  Eqs. (\ref{eqA}) and (\ref{eqB}).
Solid line: Analytical prediction given in Eq. (\ref{lam}).
}
\end{center}
\end{figure}
In a biomimetic approach, we have addressed the question of the termination of a Turing structure in a growing system. 
A free boundary is imposed at the growing part, which ensures that the wavelength of the 
pattern is not perturbed by fixed boundary conditions.
After deriving analytical expressions for the stability condition and the wavelength of the structure, we perform a systematic analysis of the effect of all dynamical parameters
on the pattern.
Apart from the variation of the diffusion coefficient of the activator, 
a well-chosen variation of the dynamical parameters leads to the desired behavior, i.e. the simultaneous loss of stability and the decrease of the wavelength.
In particular, an increase of the effective rate constant $k_{-3}R_2$, where $k_{-3}$ is the rate constant 
of the reaction injecting the inhibitor from the reservoir at the concentration $R_2$, is associated with a destabilization of the Turing pattern accompanied
by a decrease of the wavelength.\\

Imposing a spatial variation of the concentration of the reservoir R$_2$ turns out to be an appropriate protocol for chemical engineering. 
However, the proposed procedure imposes the total length of the structure but not its number of wavelengths.
In the framework of developmental biology, for example in the case of the growth of the digits or the spine of the vertebrates, the termination process 
has to respect the total number of segments for a possible variation in the length of the global structure.
Therefore, it is necessary to imagine that the system itself is able to count the number of already formed segments and to trigger the variation of a parameter leading to smaller subsequently formed segments.
If the concept of the Turing structure is kept in the formation of biological patterns, the presented results could be used to suggest such relevant parameters.
The local increase of the rate constant $k_{-3}$ that would be activated when a given number of segments has already been formed can be straightforwardly proposed.
Similarly, the local increase of the rate constant $k_2$ controlling the autocatalytic production of the activator
or the local decrease of the rate constant $k_1$ or $k_3$, associated with the absorption of the activator or the inhibitor by reservoirs, 
would lead to the desired phenomenon. The local decrease of the diffusion coefficient
of the inhibitor offers an alternative. The nature of the mechanism that would trigger such a response of the system when a given number of segments has been created remains an open question.

\section*{Acknowledgements}
This publication is part of a project that has received funding from the European Union’s Horizon 2020 research 
and innovation programme under the Marie Skłodowska-Curie grant agreement No. 711859
and has benefited from financial resources for science awarded by the Polish Ministry of Science and Higher Education in the years 2017-2021 for the implementation of an international cofinanced project.

\end{document}